\documentclass[review]{elsarticle}
\usepackage{lineno,hyperref}
\modulolinenumbers[1]
\begin{document}

%\title{Strategy for measuring cumulants of strange-hadron number distributions}
\title{Purity correction for cumulants of hyperon number distribution}

\author[mysecondaryaddress]{Toshihiro Nonaka}
\cortext[mycorrespondingauthor]{Corresponding author}
\ead{nonaka.toshihiro.ge@u.tsukuba.ac.jp}
\address{Tomonaga Center for the History of the Universe, University of Tsukuba, Tsukuba, Ibaraki, 305-0006, Japan}

\begin{abstract}
We propose a purity correction to subtract effects of combinatorial backgrounds from cumulants of hyperon number distributions.
We argue that cumulants and mix-cumulants of sidebands, whose yield is comparable with that of background particles in the signal region, 
can be used for the correction. 
The method is demonstrated in a simple toy model by introducing effects of reconstruction efficiencies and backgrounds. 
We show that topological cut parameters for hyperon reconstructions can be optimized to achieve 
the best statistical significance after purity and efficiency corrections. 
The method will enable us to measure cumulants of net-baryon, net-strangenss, and their correlations with 
better figure of merit than the conventional approach.
\end{abstract}

\maketitle

\newcommand{\ave}[1]{\ensuremath{\langle#1\rangle} }
\newcommand{\cum}[1]{\ensuremath{\langle#1\rangle}_{\rm c} }
\newcommand{\avebig}[1]{\ensuremath{\Bigl\langle#1\Bigr\rangle} }
\newcommand{\aveave}[1]{\ensuremath{\langle\!\langle#1\rangle\!\rangle} }

%\linenumbers

\section{Introduction}
Cumulants of conserved quantities like net-baryon, net-strangeness, and net-charge distributions, 
and their correlations are powerful tools to study the QCD phase diagram~\cite{nonmonotonic,Stephanov:2011pb,Koch:2005vg,Majumder:2006nq}.
These measurements have been carried out in heavy-ion collision experiments at STAR, NA61-SHINE, 
HADES, and ALICE collaborations to search for a critical point and to establish the nature of the phase transitions~\cite{Aggarwal:2010wy,net_proton,net_charge,Adamczyk:2017wsl,STAR:2021iop,STAR:2020tga,Mackowiak-Pawlowska:2021pub,Adamczewski-Musch:2020slf,ALICE:2019nbs}.
Results on the fourth-order cumulant of net-proton multiplicity distributions in Au+Au collisios hint that 
the critical point could exist at the collision energy of $\sqrt{s_{\rm NN}}\approx7.7$~GeV~\cite{STAR:2020tga,Adamczewski-Musch:2020slf}.
A negative sign of the sixth-order cumulant observed in Au+Au $\sqrt{s_{\rm NN}}=200$~GeV central collisions could suggest a smooth crossover transition at $\mu_{\rm B}=20$~MeV~\cite{STAR:2021rls,Fu:2021oaw,Friman}.

Quantitative comparisons between theory and experiments, however, are still difficult due to some 
basic difference in definitions between experimental measurements and theoretical calculations. 
One of the difference is regarding how to measure baryon numbers.
It is very difficult to measure neutrons with high efficiency in experiments, 
hence net-proton number has been measured as a proxy of net-baryon number. 
Since net-proton number is not a conserved quantity, it was proposed that the net-proton cumulants can be converted to net-baryon cumulants 
at sufficiently high collision energy at $\sqrt{s_{\rm NN}}>10$~GeV~\cite{eff_kitazawa}. 
On the other hand, there is no established way to take into account this issue at the lower energies 
where the critical point could exist. 
Furthermore, other baryons are also un-detected in experiments.
Measurements of net-proton number generally focus on the primary protons by removing secondary protons from hyperons like lambda, 
which leads to loss of baryon numbers. 
Further, the measurements of net-strangeness cumulants rely on $K^{+}$ and $K^{-}$~\cite{STAR:2020ddh}, 
while many hyperons also carry strangeness. 
It was pointed out that the loss of hyperons is crucial for the measurements of mix-cumulants between net-baryon and net-strangeness~\cite{Yang:2016xga}. 
Measuring hyperon cumulants should thus help us gain the signals of net-baryon, net-strangeness, and their correlations. 

However, this is a challenging task because of the short lifetimes of hyperons. 
They decay into daughter particles and finally lose their strangeness before hitting detectors. 
Experimentally, hyperons can be reconstructed using a invariant mass technique together with 
combinatorial backgrounds after analyzing many events and tracks. 
%A peak from the hyperons will be visible in invariant mass distributions on top of the combinatorial backgrounds after analyzing many events and tracks.
%Yields of the particles are then extracted by subtracting the backgrounds, which can be determined by fitting signal region and sidebands. 
Unfortunately, this technique is not applicable to measure higher-order cumulants of multiplicity distributions, 
since signal and background particles cannot be separated in event-by-event basis.
These measurements have been thus carried out by increasing the signal to background ratio as high as possible with optimized 
topological cut parameters for invariant mass reconstructions~\cite{STAR:2019bjj}.
On the other hand, reconstruction efficiencies tend to decrease with tightening the topological cut, which leads to 
huge statistical uncertainties of cumulants after efficiency corrections~\cite{eff_koch,eff_kitazawa,eff_xiaofeng,eff_psd_kitazawa,eff_psd_volker,Nonaka:2017kko,Luo:2018ofd}.
The measurements of net-lambdaa cumulants at the STAR collaboration are limited up to the third order~\cite{STAR:2020ddh}.

In the present paper, we propose a new method to subtract combinatorial backgrounds from cumulants of hyperons, 
which will be called "purity correction" in the rest of the paper. 
We also argue that topological cut parameters can be optimized to have the best statistical significance 
after purity and efficiency corrections through a simple toy model.

This paper is organized as follows. In Sec.~\ref{sec:sec1}, we discuss a method for the purity correction. 
The method is then applied in a simple toy model and procedures of the corrections are discussed in Sec.~\ref{sec:sec2}. 
We summarize the present study in Sec.~\ref{sec:summary}.

\section{Purity correction\label{sec:sec1}}
\subsection{Cumulants and mix-cumulants}
We consider a probability distribution function $P(m)$, where $m$ is the particle 
number at each event. 
The $r$th-order moment of $P(m)$, $\ave{m^{r}}$, is defined as:
\begin{eqnarray}
    \ave{m^{r}} &=& \left.\frac{\partial^{r}}{\partial\theta^{r}}{\rm ln}\;G(\theta)\right|_{\theta=0},\label{eq:mom1} \\
     G(\theta) &=& \sum_{m}e^{m\theta}P(m), \label{eq:mom2}
    \label{eq:mom}
\end{eqnarray}
where the bracket represents an event average and $G(\theta)$ is a moment generating function.
Similarly, the $r$th-order cumulant is defined as follows:
\begin{eqnarray}
    \ave{m^{r}}_{\rm c} &=& \left.\frac{\partial^{r}}{\partial\theta^{r}}{\rm ln}\;K(\theta)\right|_{\theta=0},\label{eq:cum1} \\
     K(\theta) &=& {\rm ln}\;G(\theta),\label{eq:cum2}
    \label{eq:cum}
\end{eqnarray}
where $K(\theta)$ is a cumulant generating function.
From Eqs.~(\ref{eq:mom1})-(\ref{eq:cum2}), cumulants up to the third-order are expressed in terms of moments as
\begin{eqnarray}
\ave{m}_{\rm c} &=& \ave{m}, \label{eq:C1}\\
\ave{m^{2}}_{\rm c} &=& \ave{m^{2}} - \ave{m}^{2}, \label{eq:C2}\\
\ave{m^{3}}_{\rm c} &=& \ave{m^{3}} - 3\ave{m^{2}}\ave{m} + 2\ave{m}^{3}. \label{eq:C3}
\end{eqnarray}
We can also consider probability distribution functions having multiple variables, whose cumulants and moments are called mix-moments and mix-cumulants, respectively. 
They are defined as follows:
\begin{eqnarray}
K(\theta_{1},...,\theta_{k}) &=& {\rm ln}\;G(\theta_{1},...,\theta_{k}), \\
G(\theta_{1},...,\theta_{k}) &=& \sum_{m_{1},...,m_{k}}\biggl(\prod_{i=1}^{k}e^{\theta_{i}m_{i}}\biggr)P(m_{1},...,m_{k}).
\end{eqnarray}
Expressions of mix-cumulants for two variables, $m_{S}$ and $m_{N}$, are obtained in terms of moments and mix-moments as 
\begin{eqnarray}
\ave{m_{S}m_{N}}_{\rm c} &=& \ave{m_{S}m_{N}} - \ave{m_{S}}\ave{m_{N}}, \\
\ave{m_{S}^{2}m_{N}}_{\rm c} &=& \ave{m_{S}^{2}m_{N}} + 2\ave{m_{S}}^{2}\ave{m_{N}} - 2\ave{m_{S}}\ave{m_{S}m_{N}} 
\nonumber \\
&&- \ave{m_{S}^{2}}\ave{m_{N}},
\end{eqnarray}

Next, let us suppose the particle number $m$ is a sum of two kinds of particles, $m=m_{S}+m_{N}$. 
Replacing $m$ in Eqs.~(\ref{eq:C1})-(\ref{eq:C3}) by $m_{S}+m_{N}$ we get
\begin{eqnarray}
\ave{m_{S}+m_{N}}_{\rm c} &=& \ave{m_{S}}_{\rm c} + \ave{m_{N}}_{\rm c}, \label{eq:sumC1}\\
\ave{(m_{S}+m_{N})^{2}}_{\rm c} &=& \ave{m_{S}^{2}}_{\rm c} + \ave{m_{N}^{2}}_{\rm c} + 2\ave{m_{S}m_{N}}_{\rm c}, 
\label{eq:sumC2}\\
\ave{(m_{S}+m_{N})^{3}}_{\rm c} &=& \ave{m_{S}^{3}}_{\rm c} + \ave{m_{N}^{3}}_{\rm c} 
+ 3\ave{m_{S}^{2}m_{N}}_{\rm c} + 3\ave{m_{S}m_{N}^{2}}_{\rm c}. \label{eq:sumC3}
\end{eqnarray}
If $m_{S}$ and $m_{N}$ are independent each other, 
the third term in the right hand side of Eq.~(\ref{eq:sumC2}) and 
the third and fourth terms in the right hand side of Eq.~(\ref{eq:sumC3}) vanish, 
which gives a famous feature of additivity of cumulants.

\subsection{Invariant mass distribution}
Experimentally, hyperons are reconstructed from their daughter particles using a invariant mass technique. 
For example, $\Lambda$(${\bar \Lambda}$) decay into $p$(${\bar p}$) and $\pi^{-}$($\pi^{+}$) 
with $63.9$\% of the branching ratio. The invariant mass is calculated for random pairs 
of $p$(${\bar p}$) and $\pi^{-}$($\pi^{+}$). 
A peak from $\Lambda$ might not be visible as it is due to huge amount of combinatorial backgrounds.
Hence, topological cuts are usually applied, e.g, on distance of the closest approach (DCA) of the daughter particles to the collision vertex, 
DCA between daughter particles, DCA of the reconstructed particle, decay length of reconstructed $\Lambda$ and so on. 
The peak in the invariant mass distribution would be visible around the mass of $\Lambda$ ($=1.1157\;{\rm GeV/c^{2}}$) 
by optimizing the topological cut parameters.
The parameters can be further tuned to have the best significance or purity.
Yields of $\Lambda$ and underlying backgrounds at a specific range of the invariant mass (which will be called "signal region") 
can be estimated by fitting the distribution around the signal peak and/or sidebands, 
and therefore the average number of $\Lambda$ is measured in the straightforward way.   
It is more difficult to calculate higher-order cumulants, since the signal and background cannot be separated in event-by-event basis.
In next subsection we derive formulas to extract true cumulants of hyperons. 

%%%%%%%%%%%%%%%%%%%%%%%%%%%%%%%%%%
\subsection{Derivation\label{sec:derivation}}
Let us start from Eq.~(\ref{eq:sumC2}). We find that the second-order cumulant of the background particles 
and mix-cumulant of the background particles are necessary to obtain the second-order cumulant for signal particles:
\begin{equation}
    \cum{m_{S}^{2}} = \cum{m_{SN}^{2}} - \cum{m_{N}^{2}} - 2\cum{m_{S}m_{N}}, 
\end{equation}
with $m_{SN}=m_{S}+m_{N}$ where $m_{S}$ and $m_{N}$ represent the number of signal and background particles in the signal region, respectively.
The last two terms in the right hand side cannot be obtained directly, so we consider to utilize sidebands of the invariant mass distribution. 
An example of a invariant mass distribution can be found in Fig.~\ref{fig:General}-(a). 
The sidebands are equally divided so the background yields at each window satisfy 
%$\ave{n_{i}}=\ave{m_{N}}$, 
$\ave{m_{R,i}}=\ave{m_{N}}$, 
where $m_{R,i}$ represents the number of background particles in the $i$th sideband window.  
If the following relations hold:
%$\cum{m_{N}^{2}}=\cum{n_{i}^{2}}$,
%$\cum{m_{S}m_{N}}=\cum{m_{S}n_{i}}$,
$\cum{m_{N}^{2}}=\cum{m_{R,i}^{2}}$,
$\cum{m_{S}m_{N}}=\cum{m_{S}m_{R,i}}$,
one obtains
\begin{equation}
    %\cum{m_{S}^{2}} = \cum{(m_{S}+m_{N})^{2}} - \cum{n_{i}^{2}} - 2\cum{m_{S}n_{i}}.\label{eq:C2_start}
    \cum{m_{S}^{2}} = \cum{m_{SN}^{2}} - \cum{m_{R}^{2}} - 2\cum{m_{S}m_{R}}.\label{eq:C2_start}
\end{equation}
Next, we consider the mix-cumulant between the signal-region and sidebands. 
This is directly measurable and decomposed as:
\begin{eqnarray}
%\cum{(m_{S}+m_{N})n_{1}} &=& \ave{(m_{S}+m_{N})n_{i}} - \ave{m_{S}+m_{N}}\ave{n_{i}} \\
%                          &=& \cum{m_{S}n_{i}} + \cum{m_{N}n_{i}} \\
%                          &=& \cum{m_{S}n_{i}} + \cum{n_{1}n_{j}},\label{eq:C2_next}
\cum{m_{SN}m_{R,i}} &=& \ave{(m_{S}+m_{N})m_{R,i}} - \ave{m_{S}+m_{N}}\ave{m_{R,i}} \\
                          &=& \cum{m_{S}m_{R,i}} + \cum{m_{N}m_{R,i}} \\
                          &=& \cum{m_{S}m_{R,i}} + \cum{m_{R,i}m_{R,j}},\label{eq:C2_next}
\end{eqnarray}
%where we utilized the relation $\cum{m_{N}n_{i}}=\cum{n_{i}n_{j}}$ with $i\neq j$.
where we utilized the relation $\cum{m_{N}m_{R,i}}=\cum{m_{R,i}m_{R,j}}$ with $i\neq j$.
From Eqs.~(\ref{eq:C2_start}) and (\ref{eq:C2_next}), one finds
\begin{eqnarray}
%\ave{m_{S}^{2}}_{\rm c} = \ave{m_{SN}^{2}}_{\rm c} - \ave{n_{i}^{2}}_{\rm c} 
%- 2\ave{m_{SN}n_{i}} + 2\ave{n_{i}n_{j}}. \label{eq:corrC2}
\ave{m_{S}^{2}}_{\rm c} = \ave{m_{SN}^{2}}_{\rm c} - \ave{m_{R,i}^{2}}_{\rm c} 
- 2\cum{m_{SN}m_{R,i}} + 2\cum{m_{R,i}m_{R,j}}. \label{eq:corrC2}
\end{eqnarray}
Similarly, one can obtain the forumula for the third-order cumulant:
\begin{eqnarray}
%\ave{m_{S}^{3}}_{\rm c} &=& \ave{m_{SN}^{3}}_{\rm c} - \ave{n_{i}^{3}}_{\rm c}
%- 3\ave{m_{SN}^{2}n_{i}} + 3\ave{n_{i}^{j}n_{2}} \nonumber \\
%&&-3 \ave{m_{SN}n_{i}^{2}} +3 \ave{n_{i}n_{j}^{2}}
%+6\ave{m_{SN}n_{i}n_{j}}_{\rm c} - 6\ave{n_{i}n_{j}n_{k}}_{\rm c}, \label{eq:corrC3}
\ave{m_{S}^{3}}_{\rm c} &=& \cum{m_{SN}^{3}} - \cum{m_{R,i}^{3}}
- 3\cum{m_{SN}^{2}m_{R,i}} + 3\cum{m_{R,i}^{2}m_{R,j}} \nonumber \\
&&-3 \cum{m_{SN}m_{R,i}^{2}} +3 \cum{m_{R,i}m_{R,j}^{2}}
+6\cum{m_{SN}m_{R,i}m_{R,j}} \nonumber \\
&&- 6\cum{m_{R,i}m_{R,j}m_{R,k}}, \label{eq:corrC3}
\end{eqnarray}
, where $m_{R,i}$, $m_{R,j}$, and $m_{R,k}$ have to be taken from different sideband windows.
The fomrmulas for mix-cumulants can be also obtained in a similar way.
See Appendinx for details. 
As we have seen in the derivations, the purity correction is justified if the probability distribution function for the background particles in the signal region is consistent with those in the sidebands in terms of certain order of cumulant. 
Most importantly, one needs to check if the width of the sideband window is properly determined so the yield is comparable with 
that of background particles in the signal region. 
Hence, it is important to estimate the background yield in the signal region as precise as possible 
by fitting the sideband, using event-mixing method, or constructing rotational backgrounds~\cite{STAR:2019bjj}. 
We also see that Eqs.~(\ref{eq:corrC2}) and (\ref{eq:corrC3}) utilize the particles from two or three sideband windows. 
One would need to prepare as many sideband windows as possible, and check the correction parameters 
as a function of invariant mass. 
Once the flatness/consitency is confirmed, one can take average over those windows to enhance the statistical accuracy for correction parameters. 
On the other hand, one naively expects that the the probability distribution function for sidebands will slightly change with increasing/decreasing the invariant mass because of the different kinematics. 
The possible variations of the correction parameters with respect to the sideband windows need to be taken into account in the systematic uncertainties of the purity correction.
%Otherwise one should use the nearest windows to the signal region, or try the other method~\cite{Arslandok:2018pcu}.

%%%%%%%%%%%%%%%%%%%%%%%%%%%%%%%%%%
\section{Test analysis\label{sec:sec2}}
In this section, we will check the validity of the purity correction through numerical simulations. 
Effects of the purity and reconstruction efficiencies will be introduced to see 
how one can achieve the best statistical significance for cumulants of the hyperon numbers in real experiments.

\begin{figure}[htbp]
    \centering
    \includegraphics[width=120mm]{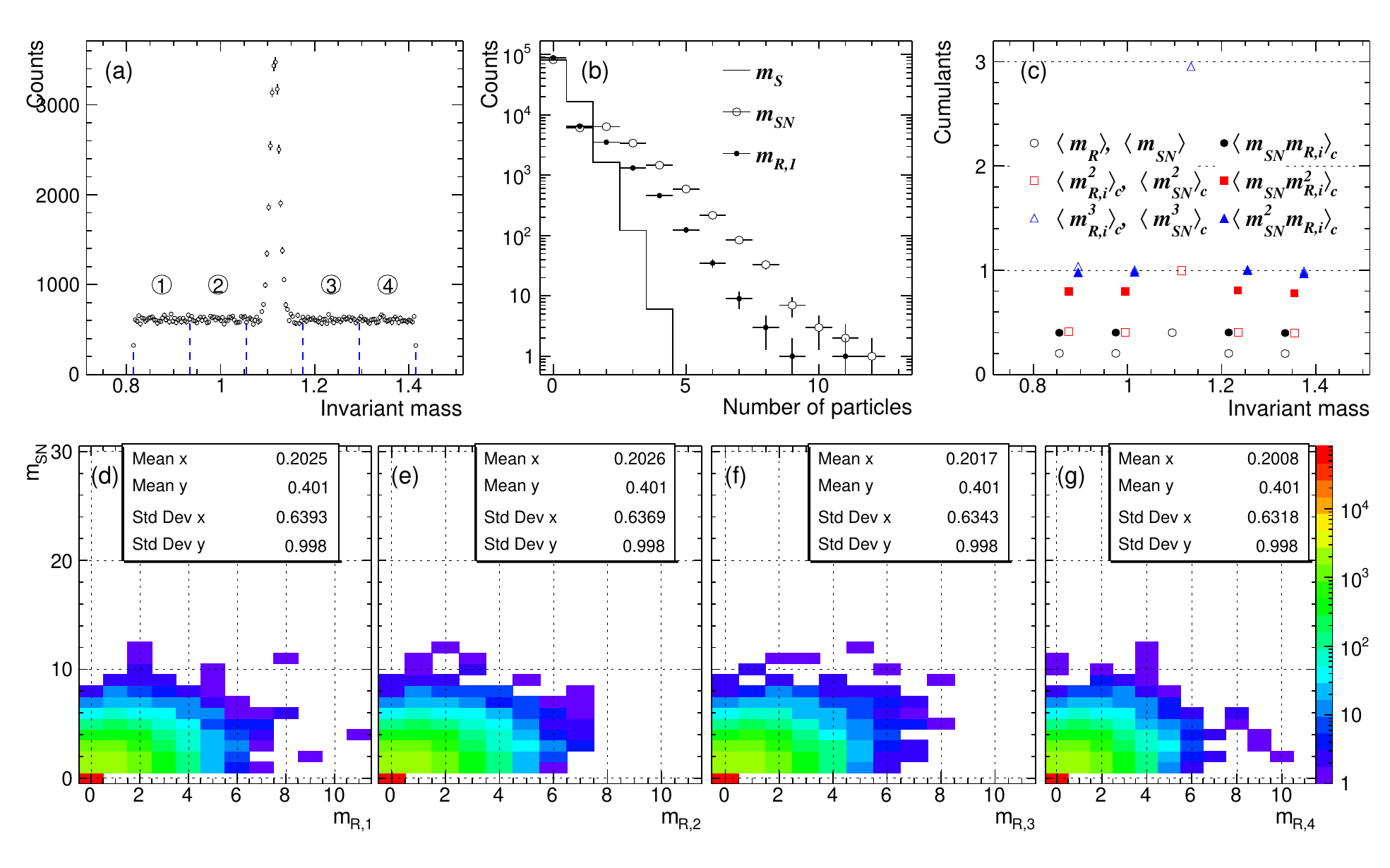}
    \caption{(a) Invariant mass distribution (b) Particle number distributions for $m_{S}$, $m_{SN}$, and $m_{SB,1}$. (c) (Mix-)Cumulants for signal candidates and sidebands' particles as a function of invariant mass. Different symbols are slightly shifted horizontally for the visibility.
    (d)-(g) Correlations between signal candidates $m_{SN}$ and particles at $i$th sideband window, $m_{R,i}$.}
    \label{fig:General}
\end{figure}

We first generate signal and background particles according to Poisson distributions. 
To implement possible correlations between signal and background particles the mean value 
for the background particles are varied event-by-event based on the signal particles 
with the purity $p=m_{S}/(m_{S}+m_{N})$.
Parameters employed in this model are summarized in Tab.~\ref{tab:param}.
For each background particle, the invariant mass is randomly allocated at $0.815<M\;(\rm GeV/c^{2})<1.415$. 
The invariant mass for signal particles is determined by a Gauss distribution whose 
mean and standard deviation are $1.115$ and $0.01$, respectively.
The resulting invariant mass distribution is shown in Fig.~\ref{fig:General}-(a). 
The distribution is divided into five windows with 
the width of $0.12\;(\rm GeV/c^{2})$. 
Boundaries for the windows are shown in blue dotted lines in Fig.~\ref{fig:General}-(a). 
The sideband windows are labeled as circled-numbers from 1 to 4.
\begin{table}[htbp]
    \centering
    \begin{tabular}{cccc} \hline
            Particles   & Distribution & Mean & Invariant mass \\ \hline\hline      
         Signal ($m_{S}$)     & Poisson & 0.2   & ($1.055,1.175$)\\ \hline
         Background ($n_{i}$, i=1,..,4) & Poisson & $m_{S}(1/p-1)$ & ($0.815,1.415$)  \\ \hline
    \end{tabular}
    \vspace{5mm}
    \caption{Parameters used in the numerical analysis. The $m_{S}$ and $n_{i}$ are the signal particle and background particles at the $i$th mass window.}
    \label{tab:param}
\end{table}
Finally, generated particles are randomly measured based on the binomial distribution: 
\begin{equation}
    B(n;\varepsilon,N)=\varepsilon^{n}(1-\varepsilon)^{N-n}\frac{N!}{n!(N-n)!},
\end{equation}
where $\varepsilon$ is a reconstruction efficiency, and $N$ and $n$ are generated and measured particles, respectively. 
In real experiments, the reconstruction efficiency tends to decrease 
with increasing the purity by tightening the topological cuts, hence 
the relation between the purity and efficiency are assumed to be anti-correlated, e.g. $\varepsilon=(1-p)^2$ for simplicity.
Particles are counted event-by-event at a signal region and four sideband windows separately, 
which will be denoted by $m_{SN}$ and $m_{R,i}$ (i=1,..,4). 
In the model we can separate signal and background particles in the signal region, 
hence they are also computed as $m_{S}$ and $m_{N}$, respectively.
 
Figure~\ref{fig:General}-(b) shows the event-by-event particle number distributions for $m_{R,1}$ and $m_{SN}$. 
A distribution for signal particles, $m_{S}$, given by the model, is also plotted. 
It is found that the distribution for $m_{SN}$ is wider and the average is larger than $m_{S}$ due to background particles.
It is not possible to measure $m_{S}$ and their cumulants directly, 
thus we consider to extract the true cumulants using those in sidebands 
according to Eqs.~(\ref{eq:corrC2}) and (\ref{eq:corrC3}). 

\begin{figure}[htbp]
    \centering
    \includegraphics[width=120mm]{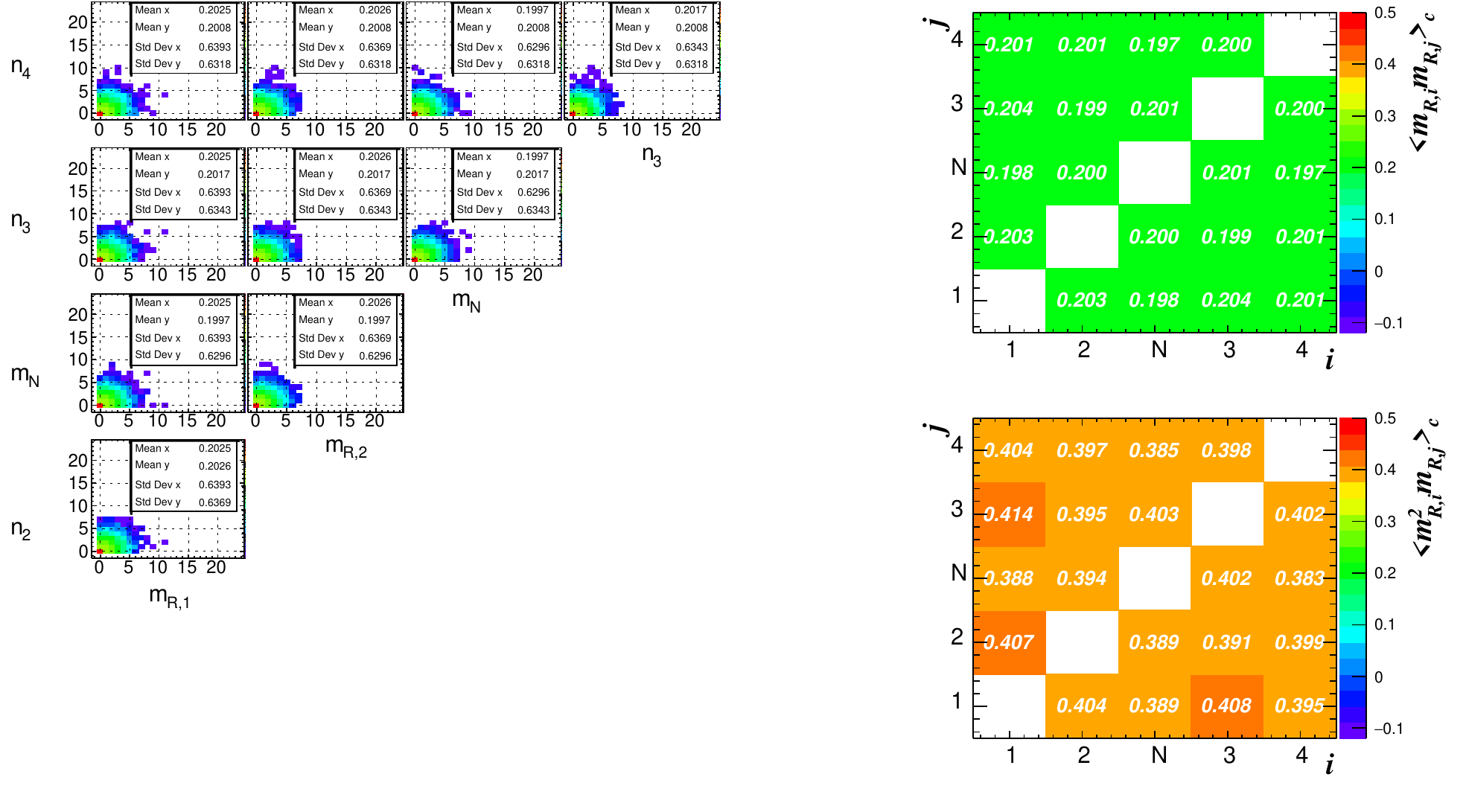}
    \caption{(Left) Correlations of particle numbers between two mass windows $i$ and $j$ ($i\neq j$). 
    (Right) The second- and third-order mix-cumulants between mass windows. 
    The values are identical between $(i,j)$ and $(j,i)$ by definition for the top panel.  }
    \label{fig:SidebandCumulant}
\end{figure}

We first calculate the cumulants for each mass window and mix-cumulants between signal region and sidebands.
Panels (d)-(g) in Fig.~\ref{fig:General} show correlations between number of signal candidates, $m_{SN}$, 
and sideband particles, $m_{R,i}$ for $i=1,..,4$. 
Sharp peaks at $(m_{R,i},m_{SN})=(0,0)$ are because that the event-by-event correlations are introduced 
between signal and background particles as shown in Tab.~\ref{tab:param}.
Cumulants up to the third-order for the sidebands and signal-region, and mix-cumulants between sidebands are plotted as a function 
of invariant mass in Fig.~\ref{fig:General}-(c). 
The cumulants at around $1.1\;(\rm GeV/c^{2})$ include both the signal and backgrounds, 
and will be substituted at the left-hand side in Eqs.~(\ref{eq:corrC2}) and (\ref{eq:corrC3}) for the purity correction. 
Next, we calculate mix-cumulants between sidebands.
The left hand side panel in Fig.~\ref{fig:SidebandCumulant} shows correlations between $m_{R,i}$, $m_{R,j}$, and $m_{N}$ ($i\neq j$). 
Corresponding second- and third-order mix-cumulants are shown in right hand side panels in Fig.~\ref{fig:SidebandCumulant}. 
Values of mix-cumulants are shown in each bin.
Averages over those bins, except for mix-cumulants including the signal region, are taken to determine 
the correction parameters, $\ave{m_{R}m_{R'}}$ in Eq.~(\ref{eq:corrC2}). 
Similarly, correlations between three mass windows such as $\ave{m_{R,i}m_{R,j}m_{R,k}}$ and $\ave{m_{SN}m_{R,i}m_{R,j}}$ 
in Eq.~(\ref{eq:corrC3}) need to be calculated for the purity correction on the third-order cumulant.


\begin{figure}[htbp]
    \centering
    \includegraphics[width=120mm]{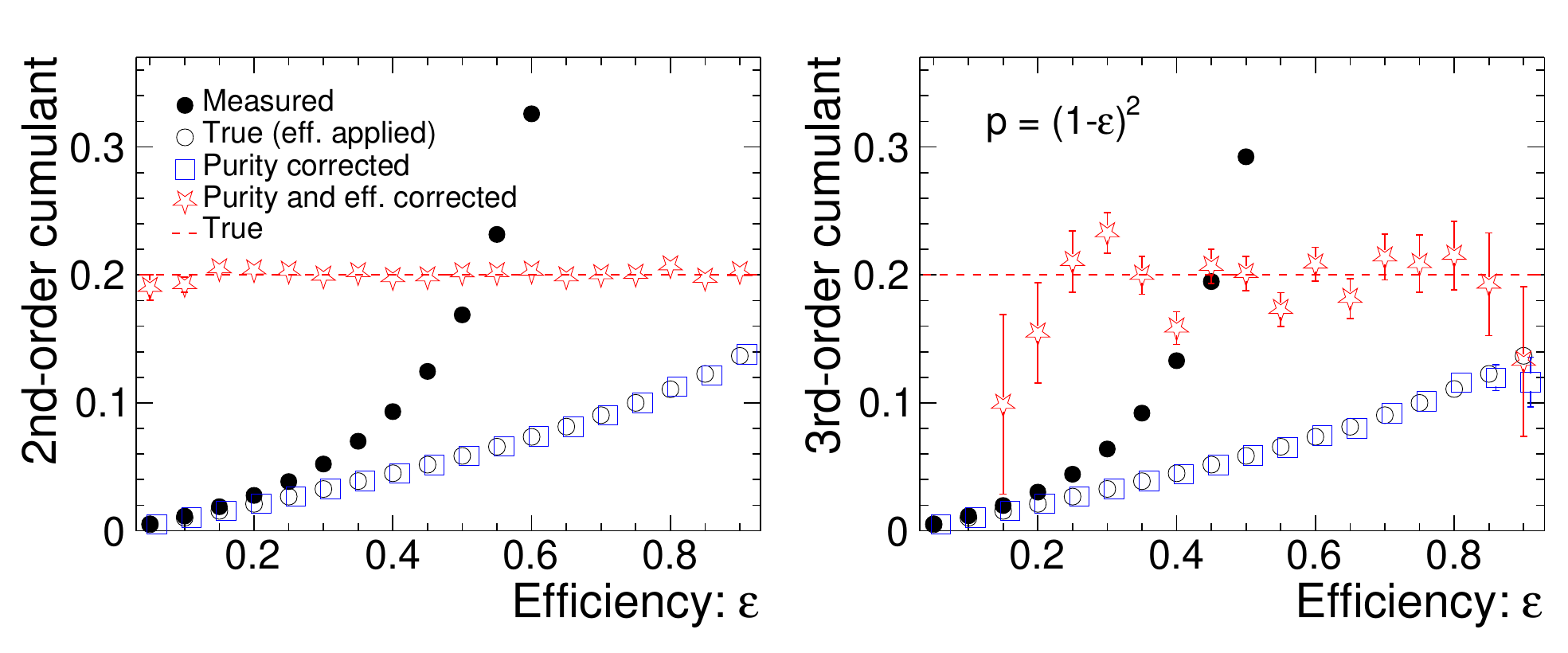}
    \caption{The second- and third-order cumulants as a function of efficiency. 
    Black filled points are for measured cumulants, black open circles are cumulants suffering the effects from the efficiency. Purity-corrrected cumulants are shown by blue squares, and fully-corrected cumulants are shown by red stars. }
    \label{fig:Result}
\end{figure}

Measured cumulants up to the third-order are shown by black solid points in Fig.~\ref{fig:Result} 
as a function of efficiency.
The results include both the effects from backgrounds and efficiencies.
The purity is varied with $p=(1-\varepsilon)^{2}$. 
The purity decreases with increasing the efficiencies, and therefore the black solid points rapidly increase with increasing the efficiency. 
We also plot the cumulant values given by the model which only incorporates the effect of the efficiency in black open circles. 
The difference between black solid and open circles thus indicates the effects from the purity given by the combinatorial backgrounds.
Next, purity corrections are applied to the black points. The purity-corrected results are shown in blue squares to be compared with the open black circles. 
It is seen that the results are consistent with the model expectation, and therefore the method we proposed does work well 
to subtract the background effects from cumulants. 

The purity-corrected cumulants are still affected by efficiencies. 
Efficiency corrections are applied to the purity-corrected cumulants according to the formulas proposed in Refs.~\cite{eff_koch,eff_kitazawa,eff_xiaofeng,eff_psd_kitazawa,eff_psd_volker,Nonaka:2017kko,Luo:2018ofd}:
\begin{eqnarray}
    Q_{2} &=& \frac{C_{2}}{\varepsilon^{2}} + \frac{C_{1}}{\varepsilon} - \frac{C_{1}}{\varepsilon^{2}}, \\
    Q_{3} &=& \frac{C_{3}}{\varepsilon^{3}} 
    + 3C_{2}\biggl(\frac{1}{\varepsilon^{2}}-\frac{1}{\varepsilon^{3}}\biggr)
    + C_{1}\biggl(\frac{1}{\varepsilon}-\frac{3}{\varepsilon^{2}}+\frac{2}{\varepsilon^{3}}\biggr), 
    %Q_{1,1} &=& \frac{C_{1,1}}{\varepsilon^{2}}, \\
    %Q_{2,1} &=& \frac{C_{2,1}}{\varepsilon^{3}} + \frac{C_{1,1}}{\varepsilon^{2}} - \frac{C_{1,1}}{\varepsilon^{3}}.
\end{eqnarray}
where $C_{r}$ and $Q_{r}$ are efficiency uncorrected and corrected $r$th-order cumulants. 
We note that in the model all particles are sampled with the same efficiency $\varepsilon$, therefore 
the formulas for the efficiency correction are in the simplest form.
The results from the full corrections including both the purity and efficiency corrections are shown in red stars. 
The fully-corrected cumulants are all found to be around the expected value of $Q_{r}$ as indicated by a dashed line at $0.2$. 

\begin{figure}[htbp]
    \centering
    \includegraphics[width=120mm]{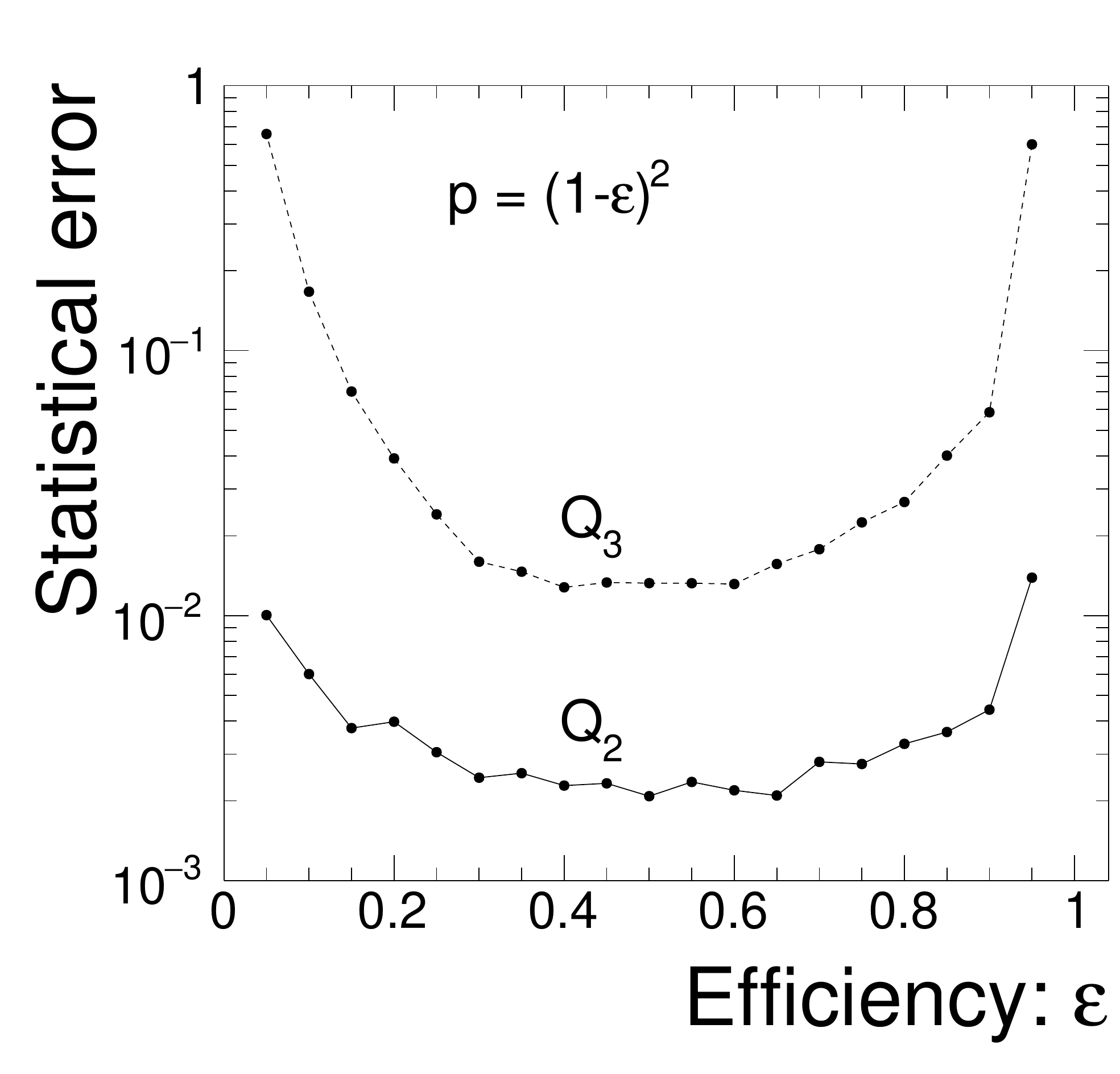}
    \caption{Statistical errors of fully-corrected cumulants as a function of efficiency.}
    \label{fig:ResultError}
\end{figure}

Now we have obtained true values of cumulants after purity and efficiency corrections,
so we can employ one result having the smallest statistical uncertainties. 
Figure~\ref{fig:ResultError} shows the statistical uncertainties of fully-corrected 
cumulants as a function of efficiency. 
It is found that we have the best statistical significance at $\varepsilon\approx0.5$ for the second- and third-order cumulants, 
and therefore the results with $\varepsilon\approx0.5$ can be employed as final results.
It should be noted that the statistical errors of cumulants  
would depend on the probability distribution function for the particles and for combinatorial backgrounds, and relation between efficiency 
and purity.


\section{Summary\label{sec:summary}}
In the present paper we proposed a new method for a purity correction 
to subtract the effect of combinatorial backgrounds from cumulants of event-by-event 
hyperon number distributions. The method is justified if the probability distribution functions for sideband windows are comparable to that for background particles in the signal region in terms of cumulants and correlations with signal candidates. 
This needs to be carefully checked in real experimental data by computing the correction parameters as a function of several sidebands and by varying the topological cut parameters. 
We have demonstrated both reconstruction efficiency and purity corrections through a toy model simulations, 
and argued that one can tune the topological cut parameters so the best statistical significance is achieved 
for final results on higher-order cumulants.
It would be also important to try the other method called "Identity Method" for an independent cross-check~\cite{Rustamov:2012bx,Ohlson:2019erm,Arslandok:2018pcu}. 

%%%%%%%%%%%%%%%%%%%%%%%%%%%%%%%%%%
\section{Acknowledgement}
TN would like to thank S. Esumi for stimulating discussions.

\bibliography{main}% Produces the bibliography via BibTeX.

\appendix
\section{Mix-cumulants}
One can derive the formulas for the purity correction on mix-cumulants. 
Let us suppose mix-cumulants between $m_{SN}=m_{S}+m_{N}$ and $n_{SN}=n_{S}+n_{N}$. 
The second-order mix-cumulant is expressed as
\begin{eqnarray}
\cum{m_{SN}n_{SN}} = \cum{m_{S}n_{S}} + \cum{m_{S}n_{N}} + \cum{m_{N}n_{S}} + \cum{m_{N}n_{N}}, \label{eq:mix_corr1} 
\end{eqnarray}
Then, we consider to express the last three terms of mix-cumulants in the right-hand side 
in terms of sideband particles, $m_{R,i}$ and $n_{R,i}$:
\begin{eqnarray}
\cum{m_{S}n_{N}} &\rightarrow& \cum{m_{S}n_{R,i}} = \cum{m_{SN}n_{R,i}} - \cum{m_{R,i}n_{R,i}}, \label{eq:mnSN}\\
\cum{m_{N}n_{S}} &\rightarrow& \cum{m_{R,i}n_{S}} = \cum{m_{R,i}n_{SN}} - \cum{m_{R,i}n_{R,i}}, \label{eq:mnNS}\\
\cum{m_{N}m_{N}} &\rightarrow& \cum{m_{R,i}n_{R,i}}. \label{eq:mnNN}
\end{eqnarray}
By substituting Eqs.~(\ref{eq:mnSN})-(\ref{eq:mnNN}) into Eq.~(\ref{eq:mix_corr1}), one obtains
\begin{equation}
    \cum{m_{S}n_{S}} = \cum{m_{SN}n_{SN}} - \cum{m_{SN}n_{R,i}} - \cum{n_{SN}m_{R,i}} + \cum{m_{R,i}n_{R,i}}.
    \label{eq:corrC11}
\end{equation}
Similarly, a correction formula for the third-order mix-cumulants is given by
\begin{eqnarray}
    \cum{m^{2}_{S}n_{S}} &=& \cum{m^{2}_{SN}n_{SN}} - \cum{m^{2}_{SN}n_{R,i}} 
    + \cum{m^{2}_{R,i}n_{R,i}} - \cum{m^{2}_{R,i}n_{SN}} \nonumber \\
    &&+ 2\cum{m_{SN}m_{R,i}n_{R,i}} - 2\cum{m_{R,i}m_{R,j}n_{R,i}}
    - 2\cum{m_{SN}n_{SN}m_{R,i}} \nonumber \\ 
    &&+ 2\cum{m_{R,i}m_{R,j}n_{SN}}
    \label{eq:corrC21}
\end{eqnarray}
with $i\neq j$.

\end{document}